\documentclass{vgtc}                          

\newif\ifnotes
\notestrue 

\ifpdf
  \pdfoutput=1\relax                   
  \usepackage{graphicx}                
  \DeclareGraphicsExtensions{.pdf,.png,.jpg,.jpeg} 
\else
  \usepackage{graphicx}                
  \DeclareGraphicsExtensions{.eps}     
\fi%

\graphicspath{{figures/}{pictures/}{images/}{./}} 

\usepackage{microtype}                 
\PassOptionsToPackage{warn}{textcomp}  
\usepackage{textcomp}                  
\usepackage{mathptmx}                  
\usepackage{times}                     
\usepackage{cite}                      
\usepackage{tabu}                      
\usepackage{booktabs}                  

\onlineid{0}

\vgtccategory{Research}

\vgtcinsertpkg

\title{The MERCADO Workshop at IEEE VIS 2023: \\ Multimodal Experiences for Remote Communication Around Data Online}

\author{Matthew Brehmer\thanks{e-mail: mbrehmer@tableau.com}\\ %
        \scriptsize Tableau Research, USA %
\and Maxime Cordeil\thanks{e-mail: m.cordeil@uq.edu.au}\\ %
     \scriptsize University of Queensland, Australia%
\and Christophe Hurter\thanks{e-mail: christophe.hurter@enac.fr}\\ %
     \parbox{1.4in}{\scriptsize \centering ENAC, University of Toulouse, France}
     \and Takayuki Itoh\thanks{e-mail: itot@is.ocha.ac.jp}\\ %
     \parbox{1.4in}{\scriptsize \centering Ochanomizu University, Japan}}

\abstract{We propose a half-day workshop at IEEE VIS 2023 on the topic of communication and collaboration around data. Specifically, we aim to gather researchers interested on multimodal, synchronous, and remote or hybrid forms of communication and collaboration within organizational and educational settings. This topic lies at the intersection of data visualization, human-computer interaction, and computer-supported collaborative work, and overlaps thematically with several prior seminars and workshops. Our intended outcomes for the workshop include assembling a corpus of inspiring examples and a design space, ideally consolidated into a survey paper, as well as the establishment of new collaborations and a shared research agenda. We anticipate a format comprised of short presentations and demos, an invited keynote or fireside chat, and a breakout group session organized around specific application domains.   
\textbf{Website}: \texttt{\href{https://sites.google.com/view/mercadoworkshop/}{sites.google.com/view/mercadoworkshop}}.
} 

\usepackage{xargs} 
\usepackage{soul}  
\usepackage{color} 
\usepackage{xspace}
\usepackage{listings}

\newcommand{\eg}{{e.g.,}\xspace}

\newcommand{\ea}{{et~al.}\xspace}

\definecolor{activegold}{RGB}{255,193,61}

\definecolor{lightorange}{rgb}{1,0.8,0.4}
\definecolor{lightorange}{RGB}{230, 170, 50}
\definecolor{lightgreen}{RGB}{121,210,121}
\definecolor{lightteal}{RGB}{121,199,210}
\definecolor{lightblue}{RGB}{100,212,239}
\definecolor{lightpurple}{RGB}{153,102,255}
\definecolor{lightred}{RGB}{245, 132, 120}
\definecolor{red}{RGB}{178,34,34}
\definecolor{gray}{RGB}{166,166,166}
\definecolor{indexBlue}{cmyk}{0.9,0.8,0,0}
\definecolor{indexGreen}{cmyk}{0.8,0.2,0.8,0.55}
\definecolor{deepblue}{cmyk}{0.9,0.75,0,0.5}
\definecolor{deepred}{cmyk}{0,0.75,0.75,0.4}
\definecolor{pink}{RGB}{214,114,0}

\usepackage{hyperref}
\usepackage{enumitem}
\setlist{topsep=0pt, leftmargin=*}

\definecolor{revision}{RGB}{255,127,14}
\definecolor{forestgreen}{RGB}{50,120,50}

\newcommand{\bdef}[1]{\vspace{1mm} \noindent{\textbf{#1}}}

\newcommandx{\todo}[1][]{
    \ifnotes
        {\textcolor{red}{[TODO] \emph{#1}}}
    \else
    \fi
}

\newcommandx{\cut}[1][]{
    \if\reviewmode
        {\textcolor{red}{\st{#1}}}
    \else
    \fi
}

\newcommandx{\sout}[1][]{
    \ifnotes
        {\textcolor{red}{\st{#1}}}
    \else
    \fi
}

\newcommandx{\claim}[1][] {\ifnotes {\textbf{\textcolor{pink}{Claim:} #1}}\else#1\fi}

\newcommandx{\maxime}[2][1=]{
    \ifnotes
        \setulcolor{lightorange}{\ul{#1}} \textcolor{lightorange}
        {[\textbf{MC:} #2]}
    \else
        #1
    \fi
}
\newcommandx{\matt}[2][1=]{
    \ifnotes
        \setulcolor{lightpurple}{\ul{#1}} \textcolor{lightpurple}
        {[\textbf{MB:} #2]}
    \else
        #1
    \fi
}

\newcommandx{\christophe}[2][1=]{
    \ifnotes
        \setulcolor{lightblue}{\ul{#1}} \textcolor{lightblue}
        {[\textbf{CH:} #2]}
    \else
        #1
    \fi
}

\newcommandx{\takayuki}[2][1=]{
    \ifnotes
        \setulcolor{lightgreen}{\ul{#1}} \textcolor{lightgreen}
        {[\textbf{TI:} #2]}
    \else
        #1
    \fi
}

\newcommandx{\revised}[1][]{
    \ifnotes
        {\textcolor{blue}{#1}}
    \else
        #1
    \fi
}

\definecolor{snapBlue}{cmyk}{0.67,1,0,0.18}
\definecolor{compGreen}{cmyk}{0.92,0,1,0.18}
\definecolor{tempOrange}{cmyk}{0,0.27,0.97,0.24}
\definecolor{tempRed}{cmyk}{0,0.69,0.91,0.32}

\begin{document}


\firstsection{Workshop Proposal: Introduction}
\label{sec:intro}

\maketitle

Throughout the last two decades, we have seen an increasing demand for remote synchronous communication and collaboration solutions across professional and educational settings. 
However, it was the COVID-19 pandemic and the urgent shift to remote and hybrid work and education that fundamentally altered how people communicate, collaborate, and teach at a distance.
Across government and enterprise organizations, people need to achieve consensus and make decisions grounded in data, and these activities now often take place via teleconference applications and collaborative productivity tools.
Meanwhile, educators at all levels have had to similarly adapt to teaching remotely, and this adaptation has presented challenges and opportunities to innovate with respect to teaching STEM (science, technology, engineering, mathematics) subjects and those that depend on numeracy, abstract representations of data, and spatial reasoning.
Popular teleconference tools such as Zoom~\cite{zoom2022}, Cisco Webex Meetings~\cite{webex2022}, Slack Huddles~\cite{slackhuddles2022}, Google Meet~\cite{meet2022}, and others typically afford multimodal communication including multi-party video and audio conferencing, screen sharing, breakout rooms, polls, reactions, and side-channel text chat functionality. 
Often these tools are used in conjunction with collaborative productivity tools, and collaboration platforms organized by channels and threads, such as Slack~\cite{slack2022} or Microsoft Teams~\cite{teams2022}, forming synchronous episodes within a larger timeline of asynchronous communication.

\begin{figure}[h!]
    \vspace{-2.5mm}
    \includegraphics[width=\linewidth]{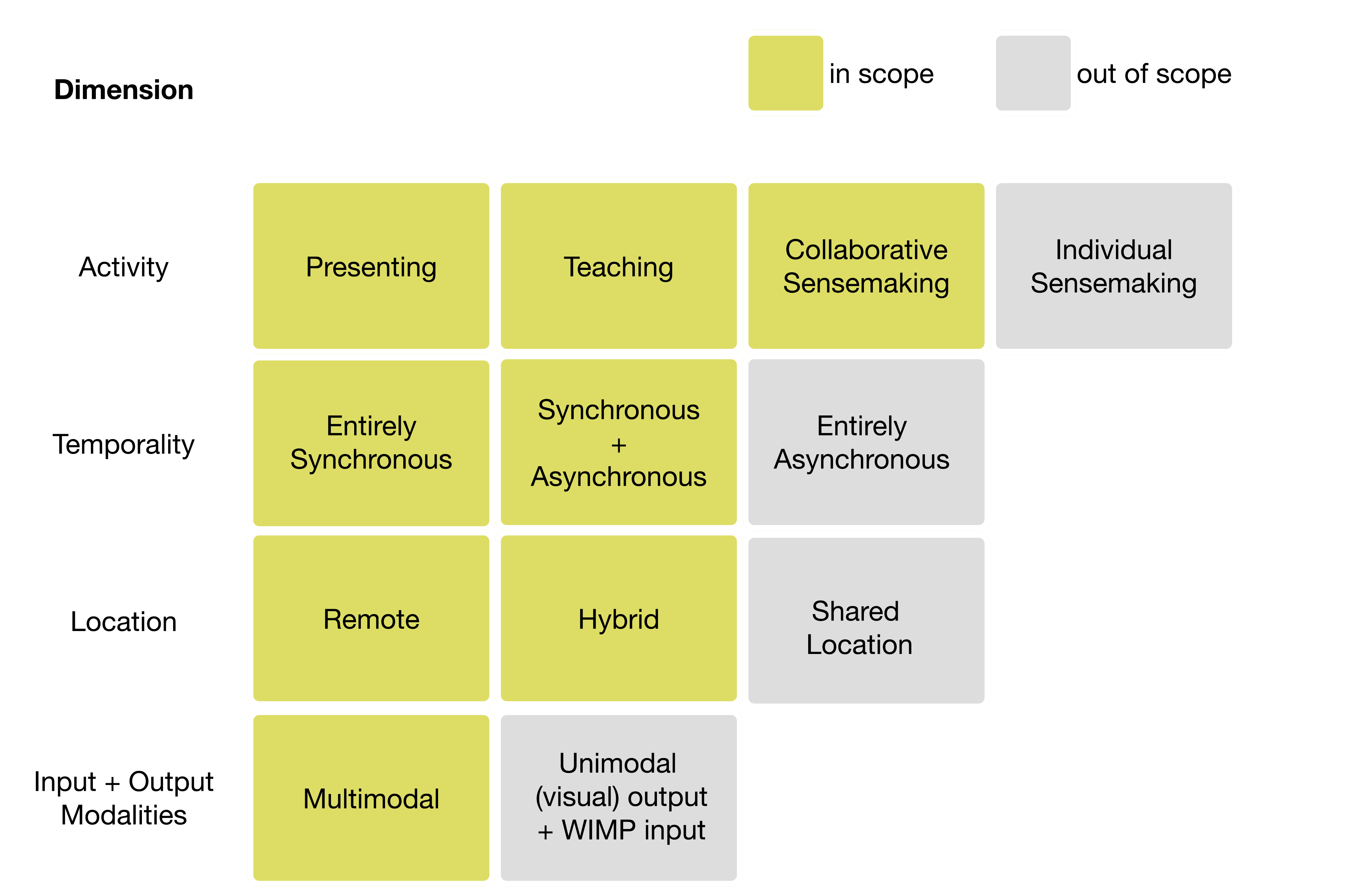}
    \vspace{-5mm}
    \caption{This diagram summarizes the scope of our proposed workshop on communication and collaboration around data.}
    \label{fig:scope}
    \vspace{-2.5mm}
\end{figure}

Despite the multimodal nature of these meetings and presentations, the presenter and audience experiences are often poor substitute for co-located communication, particularly when a speaker presents complex or dynamic multimedia content such as data visualization via screen-sharing~\cite{brehmer2021jam}. 
When screen-sharing, the presenter is often relegated to a secondary thumbnail video frame, and only they can interact with the shared content using mouse and keyboard controls. 
In contrast, consider co-located communication scenarios such as those in meeting rooms or lecture halls, where all participants can use their physical presence and body language to interact with and point to the multimedia content being discussed. 
In particular, embodied cognition research~\cite{matsumoto_nonverbal_2013} suggests that nonverbal hand gestures are essential for comprehending complex or abstract content, such in mathematics education,~\cite{aldugom_gesture_2020} in engineering and design~\cite{cash_prototyping_2016}, and in business decision-making~\cite{clarke_actions_2019}.

Some telecommunication tools have recently introduced ways to restore the missing embodied presence of a presenter as they share multimedia content such as slides, data visualization, diagrams, and interactive interfaces. 
For instance, Cisco Webex Meetings~\cite{webex2022} and Microsoft Teams~\cite{teams2022} offer functionality to segment the presenter’s outline from their webcam video and composite them in front of screen-shared content. 
Virtual camera applications have also become popular, including mmhmm\cite{mmhmm2022} and OBS Studio~\cite{obs2022}; these tools allow for considerable flexibility with respect to video compositing, and are compatible with most teleconferencing applications. 
However, only a single presenter can interact with shared content, and they must do so using standard mouse and keyboard interfaces. 

Meanwhile, advancements in extended reality (XR) and immersive analytics suggest multimodal approaches that bypass standard desktop environments. 
For instance, Flow Immersive~\cite{flow2022} is an application that allows people to present complex 3D data visualization content in mobile augmented reality (AR) or within an immersive virtual reality (VR) environment. 
More broadly, VR meeting spaces are an emerging trend in enterprise settings, in which all participants join a meeting as an avatar in an immersive 3D conference room. 
While the potential of XR for remote multimodal communication around data is promising, it also exhibits several limitations. 
The first issue is the lack of general access to affordable and comfortable hardware devices, including depth sensors and head-mounted displays; moreover, many XR applications require multi-device coordination with hand-held pointing devices or simultaneous touch-screen interaction. 
A second issue is a relatively higher amount of fatigue induced by XR applications incorporating head-mounted displays. 
A third issue is the difficulty of maintaining side-channel chat conversations in an XR environment. 
Lastly, sharing and interacting with complex and dynamic multimedia content in XR remains to be tedious and error-prone. 

Recently, an exciting alternative approach to remotely presenting rich multimedia content with remote audiences has emerged: the combination of publicly-available computer vision and speech recognition models with commodity webcam and microphones has the potential to bring the immersive experience of XR to remote communication experiences without abandoning a familiar desktop environment. 
This combination allows for real-time video compositing and background segmentation, pose and gesture recognition, and voice commands, thereby giving people multiple modalities with which to interact with representations of data.  
We have already seen applications in this space for presenting business intelligence content in enterprise scenarios~\cite{infohands2021,hall2022augmented}, for presenting STEM topics in online education\cite{liao2022realitytalk}, for personalized product marketing~\cite{liao2022realitytalk}, and for interacting with large cultural collections~\cite{rodighiero2022}, such as those in gallery and museum archives. 

Our proposed half-day workshop (\autoref{fig:scope}) lies at the crossroads of data visualization, human computer-interaction, computer-supported collaborative work, and online education, and is an opportunity to gather those who are similarly captivated by the potential of new interactive experiences for synchronous and multimodal communication and collaboration around data with remote or hybrid audiences. 

\section{Related Prior Workshops and Seminars}
\label{sec:prior}

\begin{figure}[h!]
    \vspace{-2.5mm}
    \includegraphics[width=\linewidth]{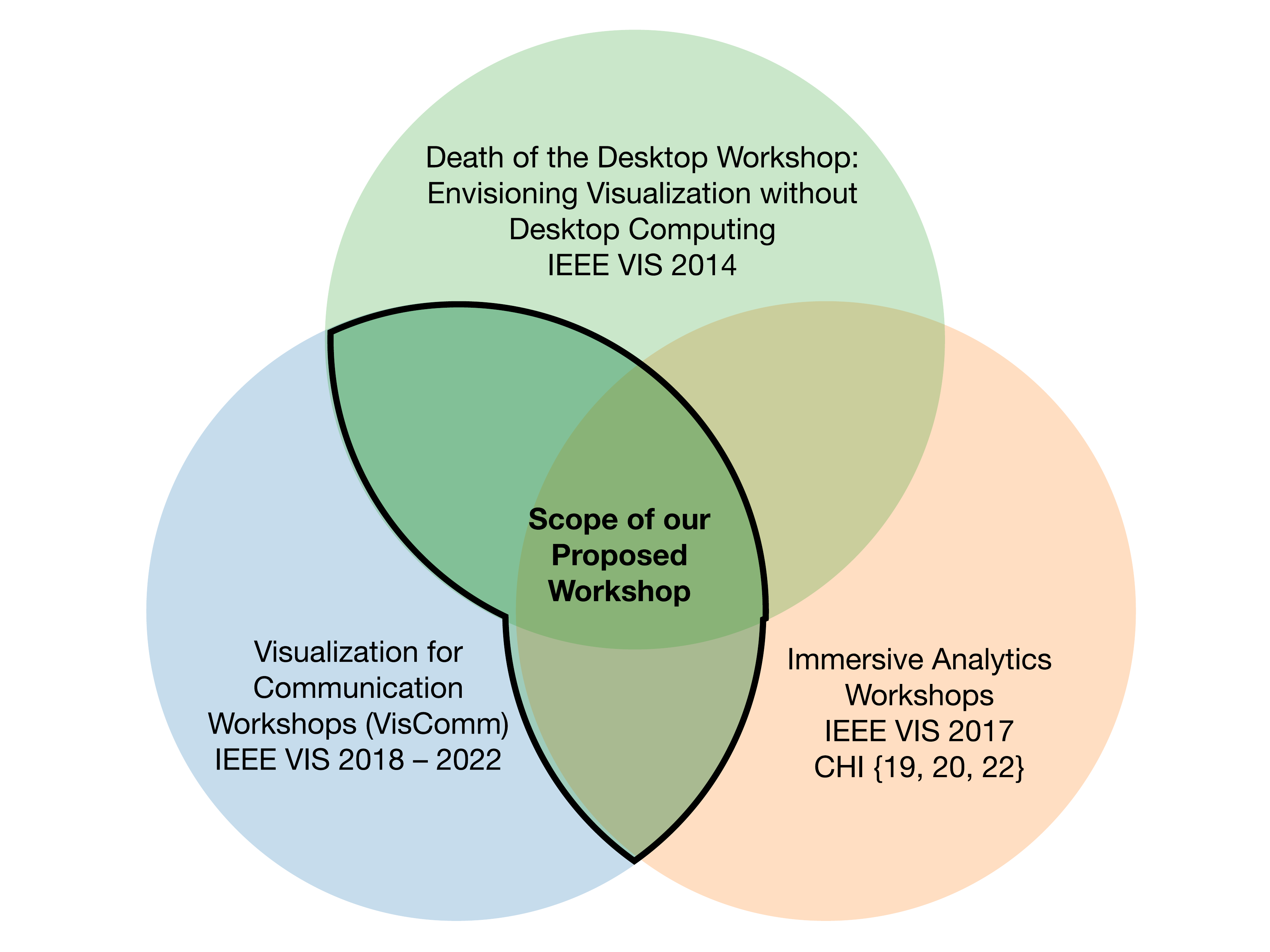}
    \vspace{-5mm}
    \caption{The area enclosed by the black contour indicates the thematic overlap between our proposed workshop and prior events.}
    \label{fig:events}
    \vspace{-2.5mm}
\end{figure}

Our proposed workshop is \textbf{not} a continuation or follow-up of a previous workshop, but it does overlap thematically with several prior workshops, as indicated in \autoref{fig:events}.
While Co-organizer M. Cordeil was a co-organizer of prior immersive analytics workshops as well as related Dagstuhl and Shonan seminars, we reiterate that the proposed workshop may involve (but will not be restricted to) immersive techniques applied to communication and collaboration. 

\section{Workshop Goals and Research Questions}
\label{sec:goals}


Our proposed workshop will focus on emerging technologies and research that can be applied to multimodal interaction for remote communication and collaboration around data. 
It will also serve as a forum to identify new application scenarios and expand upon existing ones, and as a forum to test new interaction techniques applicable across these scenarios. 
We anticipate that the results of this meeting may include contributions to several academic communities including those affiliated with the IEEE VGTC (VIS, PacificVis, ISMAR, Eurovis) and ACM SIGCHI (CHI, CSCW, UIST, ISS). 

We will invite participants to submit ideas and reflection relating to the following research questions:

\begin{itemize}
    \item Considering recent work exploring the design space of synchronous remote collaboration around data via a shared WIMP interface~\cite{neogy2020representing,schwab2020visconnect}, how can these techniques be extended to incorporate additional modalities?
    \item Considering techniques for co-located collaborative work around data, whether using conventional desktop workstations (\eg~\cite{mahyar2014clip}) or immersive augmented reality head-mounted displays (\eg~\cite{butscher2018clusters}), how can these techniques be expanded to support hybrid or remote collaboration and communication?
    \item Considering presentation techniques employed by television news broadcasters for presenting technical or data-rich stories~\cite{drucker2018} (\eg weather, finance, sports), how can these techniques be applied (while keeping production costs low) and expanded to support multi-party multimodal interaction around data at a distance?
    \item Considering presentation and video compositing techniques employed by livestreamers~\cite{chung2021,zhao2022stories} (\eg Twitch, YouTube, Facebook Live) and recorded video content creators (\eg YouTube, TikTok), how can these techniques be applied during synchronous communication and collaboration around data, as well as in conjunction with multimodal interaction (\eg pose / gesture input, voice prompts, proxemic interaction)?
    \item Considering the techniques by which individuals display and interact with representations of data in XR (AR / VR, \eg~\cite{chen2019,lee2020data}) how can we extend or adapt these techniques? In other words, how can techniques initially designed with expensive or exclusive hardware be adapted to low-cost, accessible commodity input and output devices? Similarly, how could augmented video techniques designed for use with depth sensors~\cite{saquib_interactive_2019} and pointing devices~\cite{Perlin2017} be similarly adapted?
    \item Currently, most teleconference applications assume a single speaker / presenter, with other participants in audience roles. How can we support multi-party augmented video interaction, such as with Gr{\o}nb{\ae}k~\ea's MirrorBlender project~\cite{gronbaek2021mirrorblender}? Alternatively, how can we support both \textit{`sage on the stage'} and \textit{`guide on the side'} style communication~\cite{king1993sage}, differentiating an orator from a discussion facilitator. Similarly, how can we support formal, linear, and scripted presentations as well as informal, unscripted, interruption-prone, and collaborative discussions? Using Brehmer and Kosara's musical performance analogy~\cite{brehmer2021jam}, the former experience is likened to a \textit{`concert recital} while the latter one is likened to a \textit{`jam sessions'}~\cite{brehmer2021jam}).
    \item Overall, what are the dimensions of the design space for multimodal and synchronous communication and collaboration around data? Where does existing work fit within this design space and which parts remain underexplored?    
\end{itemize}

Ultimately, our aim is to identify a timely and urgent research agenda that the community gathered during this workshop can pursue. 
We aim to report the emerging research directions that we will identify in a top quality outlet research venue such as IEEE TVCG or ACM CHI. 

\section{Format and Planned Activities}
\label{sec:format}

As a half-day workshop, we anticipate two 75-minute workshop sessions on either side of a 30-minute coffee break, assuming a schedule similar to recent VIS conferences.
The two sessions will consist of:

\begin{itemize}
    \item 
    \textbf{Session 1}: a general introduction to the workshop aims and activities (15 mins), and 60 mins allocated for short papers presentations and active discussions (elicitation of emerging research topics, usage scenarios, and a shared research agenda). 
    \item
    \textbf{Session 2}: keynote or fireside chat (30 min), breakout groups to ideate on the design space across application domains (30 mins), groups report back and wrap up (15 mins in the second session).
\end{itemize}



Given our topic, we are open to hosting a remote keynote speaker or fireside chat guest. 
However, as VIS 2023 will be held in Australia, we see this workshop as an opportunity to invite a local practitioner to inspire attendees.

\section{Accommodations and Contingencies}
\label{sec:accommodations}

To run the workshop we require a standard VIS workshop room with AV (microphone for presenter and microphone for questions in the audience, projector). If the workshop requires some hybrid form of participation, participants will join with their laptop (muted), and presenters use the conference/workshop org laptop to present on zoom and projector. For online collaboration tool, a Miro\footnote{\texttt{miro.com}} page created with a pro account so anyone with the link can join without having to sign in would be required. Talks can be recorded with zoom if needed.



If VIS 2023 will run in hybrid mode, we would gladly host this workshop concurrently in person and via video-conference, allowing remote presenters and keynote speakers to participate in the activities.
However, given the theme of remote / hybrid collaboration of this workshop, we would also request hybrid participation regardless of the overall conference format. 
If this workshop proposal is accepted, we would gladly initiate conversations with the VIS organizing committee at an early stage to ensure a successful hybrid event.

\section{Publication Plan}
\label{sec:publication}


Those who want to participate will submit a short paper (two to four pages) in the style of short provocation or reflection statements, case studies, or work-in-progress reports. 
We will accept submissions using the IEEE VGTC Conference Style Template format, a pictorial format similar to submissions made to the VIS Arts Program\footnote{\texttt{visap.net}}, or in an interactive article format similar to submissions to the VISxAI workshop\footnote{\texttt{visxai.io}}. 
We will also encourage the submission of accompanying videos. We will host accepted submissions on the workshop's website (\texttt{\href{https://sites.google.com/view/mercadoworkshop/}{sites.google.com/view/mercadoworkshop}}.), however will we also encourage authors to archive their publications using open access services. The accepted submissions will not be considered archival but could be cited, and we will encourage reuse of the content in a follow-up publications. Submissions will not be anonymous and authors must include the full name of authors, their emails, and affiliations. Lastly, at least one author for an accepted paper will need to register for the VIS conference. 

\section{Call for Participation and Timeline}
\label{sec:participation}




We plan to announce our call for participation around the time of the CHI and PacificVis conferences (mid-April), advertising on IEEE and ACM mailing lists as well as on social media. 
Assuming a camera-ready deadline of August 21, 2023, we plan to have a submission deadline following the first-round notification of full VIS papers (mid-June), with a short review period concluding with author notifications in mid-July.   
In addition to workshop papers, we will also invite short provocation, pictorial, and demo submissions with a later deadline in early September.

Our main selection criteria is the relevance of the submission to the topic of novel techniques for communicating and collaborating around data with remote and hybrid audiences. 

\section{Organizers}
\label{sec:organizers}


\bdef{Matthew Brehmer} is a senior research staff member of Tableau Research in Seattle, USA, where he specializes in new experiences for interpersonal communication with and around data (\eg~\cite{brehmer2021jam,infohands2021,hall2022augmented}). Prior to joining Tableau, he was a postdoctoral researcher at Microsoft Research specializing in storytelling and visualization beyond the desktop, which followed his PhD research on information visualization at the University of British Columbia. He was a co-organizer of the first workshop on visualization on mobile devices at CHI 2018~\cite{lee2018data}, and a co-organizer of the VisInPractice~\cite{visinpractice} event at IEEE VIS between 2018 and 2021. In 2022, he was elected to the VIS Executive Committee (VEC) and appointed to the IEEE Visualization and Computer Graphics Technical Community (VGTC) Executive Committee. \textbf{Website}: \texttt{\href{https://mattbrehmer.ca/}{mattbrehmer.ca}}.

\bdef{Maxime Cordeil} is a Senior Lecturer at the University of Queensland, Australia. Dr. Cordeil has been recognised Australia’s top researcher in computer graphics (2021, 2022). His research focuses on human-computer interaction, data visualisation and analytics. He has published over 60 journal and conference in top venues such as ACM CHI, IEEE VIS or IEEE VR. Dr. Cordeil is a key international member of the Immersive Analytics community of researchers, and has organised several workshops on the topic of Immersive Analytics (``IA Workshop series'' at VIS 2017, CHI 2018, CHI 2019, CHI 2020, and CHI 2022). The activities of the IA community focuses on designing and evaluating the future graphical user interfaces for data analysis in Virtual / Augmented Reality. \textbf{Website}: \texttt{\href{https://sites.google.com/view/cordeil/home}{sites.google.com/view/cordeil}}.

\bdef{Christophe Hurter} is a Professor working at the University of Toulouse, France, leading the Interactive Data Visualization group (DataVis) of the French Civil Aviation University (ENAC). His research covers explainable A.I. (XAI), big data manipulation and visualization (InfoVis), immersive analytics, and human-computer interaction (HCI). He investigates the design of scalable visual interfaces and the development of pixel-based techniques. He is an associate researcher at the research center for the French Military Air
Force Test Center (CReA, Base militaire de Salon de Provence) and at the Brain and Cognition Research Center (CerCo, Hospital University Center of Toulouse). He published 2 books, 4 book chapters, 20 patents, 25 journal papers, more than 100 per reviewed international research papers. \textbf{Website}: \texttt{\href{http://recherche.enac.fr/~hurter/}{recherche.enac.fr/~hurter}}.

\bdef{Takayuki Itoh} is a full professor of the department of information sciences in Ochanomizu University, Japan since 2011, and the director of the center for artificial intelligence and data science of the university since 2019. 
He was a researcher at Tokyo Research Laboratory of IBM Japan during 1992 to 2005. 
He has been an associate professor in Ochanomizu University since 2005, and a full professor since 2011. 
He is the general chair of Graph Drawing 2022, 
the general chair of IEEE Pacific Visualization 2018, 
short paper co-chair of IEEE VIS 2023,
and organizing members of other many international conferences.
His representative studies include fast isosurface generation \cite{itohTVCG1995, itohTVCG2001}, hierarchical data visualization \cite{itohTVCG2004, itohCGA2006}, network visualization \cite{itohPVis2009, itohCGA2015, itohTVCG2020} and multidimensional data visualization \cite{itohJVLC2017}. \textbf{Website}: \texttt{\href{http://itolab.is.ocha.ac.jp/~itot/}{itolab.is.ocha.ac.jp/~itot}}.

\bibliographystyle{abbrv-doi}

\bibliography{main}
\end{document}